\documentclass[prb,eqsecnum,aps]{revtex4}
\usepackage{graphicx}
\newcommand{\e}{{\rm e}}
\newcommand{\ep}{\varepsilon}

\newcommand{\la}{\label} 

\newcommand{\be}{\begin{equation}}
\newcommand{\ee}{\end{equation}}
\newcommand{\ba}{\begin{eqnarray}}
\newcommand{\ea}{\end{eqnarray}}

\begin{document}
\title{The Complete Characterization of Fourth-Order Symplectic Integrators
with Extended-Linear Coefficients}

\author{Siu A. Chin}

\affiliation{Department of Physics, Texas A\&M University,
College Station, TX 77843, USA}

\begin{abstract}
The structure of symplectic integrators up to fourth-order 
can be completely and analytical understood when the factorization 
(split) coefficents are related linearly but with a uniform 
nonlinear proportional factor. 
The analytic form of these {\it extended-linear} symplectic 
integrators greatly simplified
proofs of their general properties and allowed easy 
construction of both forward and non-forward fourth-order algorithms 
with arbitrary number of operators. Most fourth-order
forward integrators can now be derived analytically
from this extended-linear formulation without the use of
symbolic algebra.  

\end{abstract}
\maketitle

\section {Introduction}

Evolution equations of the form
\be
w(t+\ep)={\rm e}^{\ep(T+V)} w(t),
\label{two}
\ee
where $T$ and $V$ are non-commuting operators,
are fundamental to all fields of physics ranging from
classical mechanics\cite{yoshi,hairer,mcl02,chinchen03,chinsante}, 
electrodynamics\cite{hirono,lee},
statistical mechanics\cite{ti,jang,fchinl,chincor} to quantum 
mechanics\cite{feit,chinchen01,chinchen02}.
All can be solved by approximating
${\rm e}^{\ep(T+V)}$ to the $(n+1)$th order in the product form
\be
{\rm e}^{\ep (T+V )}=\prod_{i=1}^N
{\rm e}^{t_i\ep T}{\rm e}^{v_i\ep V}+O(\ep^{n+1})
\label{prod}
\ee
via a well chosen set of factorization (or split) coefficients 
$\{t_i\}$ and $\{v_i\}$. The resulting algorithm is then
$n$th order because the algorithm's Hamiltonian is 
$T+V+O(\ep^n)$. By understanding this single approximation,
computational problems in diverse fields of physics can all 
be solved by applying the same algorithm.  

Classically, (\ref{prod}) results in a class of composed or factorized 
symplectic integrators. While the conditions on $\{t_i\}$ and $\{v_i\}$ 
for producing an {\it n}th order algorithm can be stated, these 
{\it order conditions} are highly nonlinear and analytically opaque.
In many cases\cite{chinchen02,kos96,ome02,ome03}, elaborate symbolic 
mathematical programs are needed to produce even fairly low order algorithms 
if $N$ is large.  In this work, we show that the structure of most 
fourth-order algorithms, including nearly all known forward ($\{t_i,v_i\}>0$) 
integrators, can be understood and derived on the basis that $\{v_i\}$ 
and $\{t_i\}$ are linearly related but with a uniform nonlinear
proportional factor. 
This class of {\it extended-linear integrator} is sufficiently complex 
to be respresentative 
of symplectic algorithms in general, but its transparent structure makes 
it invaluable for constructing integrators up to the fourth-order. 
In this work we prove three important theorems on the basis of which,
many families of fourth-order algorithms can be derived with analytically
known coefficients, including all known forward integrators up to $N=4$.

\section {The Error Coefficients}

The product form (\ref{prod}) has the general expansion
\ba
&&\prod_{i=1}^N
\e^{t_i\ep T}\e^{v_i\ep V}
=\exp\biggl( \ep e_T T+\ep e_V V+\ep^2 e_{TV}[T,V]\nonumber\\
&&+\,\ep^3 e_{TTV}[T,[T,V]]+\ep^3 e_{VTV}[V,[T,V]]+\cdots\biggr).
\label{prodform} 
\ea 
We have previously\cite{nosix} described in detail how the error coefficients
$e_T$, $e_V$, $e_{TV}$, $e_{TTV}$, and $e_{VTV}$ can be
determined from $\{t_i\}$ and $\{v_i\}$:
\begin{equation}
e_T=\sum_{i=1}^N t_i\,,\quad\quad e_V=\sum_{i=1}^N v_i\,,
\label{tvcon}
\end{equation}
\be {1\over 2}+ 
e_{TV}=\sum_{i=1}^N\nabla\! s_i u_i\,,
\label{etv} 
\ee  
\ba &&{1\over 3!}+{1\over
2}e_{TV}+e_{TTV}={1\over
2}\sum_{i=1}^N\nabla\!s_i^2 u_i\,,
\label{ettv}\\
&&{1\over 3!}+{1\over 2}e_{TV}-e_{VTV}
={1\over 2}\sum_{i=1}^N\nabla\!s_i u_i^2\,,
\label{evtv}
\ea
where we have defined useful variables
\be s_i=\sum_{j=1}^i t_j\,,\quad u_i=\sum_{j=i}^N v_j\,, 
 \label{si} 
\ee 
and the backward finite differences 
\be
\nabla\!s_i^n=s_i^n-s_{i-1}^n,
\label{nab}
\ee
with property
\be
\sum_{i=1}^{N}\nabla\!s_i^n=s_N^n\,(=e_T^n=1).
\label{sumn}
\ee
We will always assume that the primary constraint
$e_T=1$ and $e_V=1$ are satisfied so that (\ref{sumn}) sums to
unity. Satisfying these two primary constriants is sufficient to
produce a first order algorithm. For a second-order algorithm, 
one must additionally forces $e_{TV}=0$. For a third-order algorithm, 
one further requires that $e_{TTV}=0$ and $e_{VTV}=0$. 
For a fourth-order algorithm, one has to satisfy the third-order 
constraints with coefficients $t_i$ that are left-right symmetric.
(The symmetry for $v_i$ will follow and need not be imposed {\it a priori}.)
Once the primary conditions $e_T=1$ and $e_V=1$ are imposed,
the constraints equations (\ref{etv})-(\ref{evtv}) are highly
nonlinear and difficult to decipher analytically. In this work,
we will show that (\ref{etv}) can be satisfied for all $N$ 
by having $\{v_i\}$ linearly related to $\{t_i\}$ (or vice versa). 
The coefficients $e_{TTV}$ and $e_{VTV}$ can then be evaluated simply 
in terms of $\{t_i\}$ (or $\{v_i\})$ alone. 
This then completely determines the structure of third and fourth-order 
algorithms.
   
\section {The Extended-Linear Formulation }

The constraint $e_{TV}=0$ is satisfied if
\be 
\sum_{i=1}^N\nabla\! s_i u_i={1\over 2}.
\label{evtc} 
\ee  
If we view $\{t_i\}$ as given, this is a linear equation for
$\{u_i\}$. Knowing (\ref{sumn}), a general solution for $u_i$ 
in terms of $s_i$ and $s_{i-1}$ is
\be
u_i=\sum_{n=1}^M C_n{{\nabla\! s^n_i}\over{\nabla\! s_i}},
\quad {\rm with}\quad \sum_{n=1}^M C_n=\frac12.
\label{uexp}
\ee
The coefficients $C_n$ respresent the intrinsic freedom in $\{v_i\}$ to
satisfy any constraint as expressed through its relationship to $\{t_i\}$. 
The expansion (\ref{uexp}) is in increasing powers of $s_i$ and $s_{i-1}$. If we
truncated the expansion at $M=2$, then for $i\ne 1$, $u_i$ is linearly 
related to $\{s_i\}$, {\it i.e.},
\be
u_i=C_1+C_2{{\nabla\! s^2_i}\over{\nabla\! s_i}}=C_1+C_2(s_i+s_{i-1}).
\la{ui}
\ee
For $i=1$, since we must satisfy the primary constraint $e_V=1$, 
we must have
\be
u_1=1.
\la{uone}
\ee
In this case, the constriant (\ref{evtc}) takes the form
\be 
\sum_{i=1}^N\nabla\! s_i u_i
=t_1+C_1(1-t_1)+C_2(1-t_1^2)=\frac12.
\label{evp} 
\ee 
The complication introduced by $u_1=1$, in this, and in other similar sums,
can be avoided without any loss of generality by decreeing  
\be
t_1=0,
\la{tsetone}
\ee
so that (\ref{evp}) remains
\be
C_1+C_2=\frac12.
\la{tv2}
\ee
For $i\ne 1\ne N$, (\ref{ui}) implies that
\be
v_i=-C_2(t_i+t_{i+1}).
\la{vi}
\ee
Since $v_1=u_1-u_2=1-C_1-C_2 t_2$, by virtue of (\ref{tv2}),
\be
v_1=\frac12+C_2(1-t_2).
\la{vonea}
\ee
Similarly, since $v_N=u_N=C_1+C_2(2-t_N)$, we also have
\be
v_N=\frac12+C_2(1-t_N).
\la{vn}
\ee
Given $\{t_i\}$ such that $t_1=0$, the set of $\{v_i\}$ defined by
(\ref{vi})-(\ref{vn}) automatically satisfies $e_V=1$ and $e_{TV}=0$.
If $C_2$ were a real constant, then $\{v_i\}$ is linearly related to 
$\{t_i\}$. However, in most cases $C_2$ will be a function of
$\{t_i\}$ and the actual dependence is nonlinear. 
But the nonlinearity is restricted to $C_2$, which is
the same for all $v_i$. We will call this special form of 
dependence of $v_i$ on $\{t_i\}$, {\it extended-linear}. 
For a given set of $t_i$, 
(\ref{vi})-(\ref{vn}) defines our class of 
extended-linear integrators with one remaining
parameter $C_2$.

For extended-linear integrators as described above, 
one can easily check 
that the sums in (\ref{ettv}) and (\ref{evtv}) can be evaluated as
\ba 
&&\sum_{i=1}^N\nabla\!s_i^2 u_i=C_1+C_2+gC_2=\frac12+g C_2,\qquad\quad
\la{fttv}\\
&&\sum_{i=1}^N \nabla\!s_i u_i^2=(C_1+C_2)^2+gC_2^2=\frac14+g C_2^2.
\la{fvtv}
\ea
Again the complication introduced by $u_1=1$ is avoided by
decreeing $t_1=0$.
The quantity $g$ is a frequently occuring sum defined via
\be
\sum_{i=1}^N {{\nabla\!s_i^2\nabla\!s_i^2}\over{\nabla\!s_i}}=1+g,
\label{gdef}
\ee
with explicit form
\be
g=\sum_{i=1}^N s_is_{i-1}(s_i-s_{i-1})={1\over 3}(1-\delta g),
\label{gsec}
\ee
where
\be
\delta g=\sum_{i=1}^N t_i^3.
\label{delg}
\ee
Much of the mechanics of dealing with these sums have been worked 
out in Ref.\cite{nosix}. However, their use and interpretation here are 
very different. From (\ref{ettv}) and (\ref{evtv}), we have
\ba && e_{TTV}=\frac1{12}+\frac12 g C_2,
\label{ttvf}\\
&& e_{VTV}=\frac1{24}-\frac12 g C_2^2.
\label{vtvf}
\ea
Both are now only functions of $\{t_i\}$ through $g$. 

\section {Fundamental Theorems  }

We can now prove a number of important results:

\noindent{\bf Theorem 1.} For the class of extended-linear
symplectic integrators defined by $t_1=0$ and (\ref{vi})-(\ref{vn}),
if $\{t_i\}>0$ for $i\ne1$ such that
$e_T=1$, then $e_{TTV}\ne e_{VTV}$.

\noindent
{\it Proof:} Setting $e_{TTV}= e_{VTV}$	produces a quadratic
equation for $C_2$,
\be
C_2^2+C_2+\frac{1}{12g}=0
\ee
whose discriminant
\be
D=b^2-4ac=-\frac{\delta g}{1-\delta g}
\ee
is strictly negative (since if $e_T=1$, then
$1>\delta g>0$). Hence no real solution exists for $C_2$.
This is a fundamental theorem about positive-coefficient 
factorizations. This was first proved generally in the context of 
symplectic corrector (or process) algorithms by
by Chin\cite{chincor} and by Blanes and Casas\cite{blanes05}. 
If $e_{TTV}$ can never equal
$e_{VTV}$, then no second order algorithm with positive coefficients
can be corrected beyond second order with the use of a corrector.

As a corollary, for $\{t_{i>1}\}>0$, $e_{TTV}$ and $e_{VTV}$ cannot 
both vanish. This is the content of the Sheng-Suzuki 
Theorem\cite{sheng,suzukinogo}:
there are no integrators of order greater than two of the
form (\ref{prodform}) with only positive
factorization coefficients. Our proof here is restricted to extended-linear
integrators, but can be interpreted more generally as it is done in
Ref.\cite{nosix}. Blanes and Casas\cite{blanes05} have also given a
elementary proof of this using a very weak {\it necessary}
condition. Here, for extended-linear integrators, we can be very precise in 
stating how both $e_{TTV}$ and $e_{VTV}$ fail to vanish. 
We have, from (\ref{ttvf}), if $e_{TTV}=0$, then
\be
 C_2 = -\frac1{2(1-\delta g)},
\quad e_{VTV}=-{1\over {24}}{{\delta g}\over{(1-\delta g})}.
\la{vtvv}
\ee
Similarly, from (\ref{vtvf}), if $e_{VTV}=0$, then
\be 
C_2 = -\frac1{2\sqrt{1-\delta g}},
\quad e_{TTV} =\frac1{12}\left (1-\sqrt{1-\delta g}\right ). 
\la{ttvv}
\ee
Satisfying either condition forces $C_2$ to be a
function of $\{t_i\}$ through $\delta g$.
From Ref.\cite{nosix}, we have learned that the value given by (\ref{vtvv})
is actually an upperbound for $e_{VTV}$ if $\{t_{i>1}\}>0$ and
$e_{TTV}=0$.
Similarly, in general, the value given by (\ref{ttvv}) is 
a lower bound for $e_{TTV}$ if $\{t_{i>1}\}>0$ and
$e_{VTV}=0$. Our class of extended-linear integrators are all algorithms
that attain these bounds for positive $t_{i>1}$. Note that in
(\ref{ttvv}) we have discarded the positive solution for $C_2$
which would have led to negative values for the $v_i$ coefficients.

For the study of forward integrators where one requires $\{t_{i>1}\}>0$, it
is useful to state (\ref{vtvv}) as a theorem:

\noindent{\bf Theorem 2a.} For the class of extended-linear
symplectic integrators defined by (\ref{vi})-(\ref{vn})
with $t_1=0$, $e_T=1$, and 
$C_2$, $e_{VTV}$ given by
\be
C_2=-\frac1{2\phi},\quad
e_{VTV}=-\frac1{24}\Bigl(\frac1\phi-1\Bigr),\quad\phi=1-\delta g,
\la{cande}
\ee
one has
\be
\prod_{i=1}^N
\e^{t_i\ep T}\e^{v_i\ep V}
=\exp\biggl(\ep (T+V)+\,\ep^3e_{VTV} [V,[T,V]]+\cdots\biggr).
\label{th2a} 
\ee
For $t_1=0$, the first operator $\e^{v_1\ep V}$ classically
updates the velocity (momentum) variable.
Theorem 2a completely described the structure of these
{\it velocity}-type algorithms.

If one now interchanges 
$T\leftrightarrow V$ and $\{t_i\}\leftrightarrow \{v_i\}$, then
$[T,[T,V]]$ transforms into $[V,[T,V]]$ with a sign change.
Hence, one needs to interpret $e_{TTV}$ in (\ref{ttvv}) as $-e_{VTV}$,
yielding:  

\noindent{\bf Theorem 2b.} For the class of extended-linear
symplectic integrators defined by
\be
t_1=\frac12+C_2(1-v_2),\quad t_N=\frac12+C_2(1-v_N),\quad
t_i=-C_2(v_i+v_{i+1}),
\la{ti}
\ee 
with $v_1=0$, $e_V=1$, and
\be
C_2 = -\frac1{2\phi^\prime},
\quad\quad e_{VTV} =-\frac1{12}(1-\phi^\prime ), 
\la{candep}
\ee
where
\be
\phi^\prime=\sqrt{1-\delta g^\prime},\qquad \delta g^\prime=\sum_{i=1}^N v_i^3,
\ee
one has
\be
\prod_{i=1}^N \e^{v_i\ep V}
\e^{t_i\ep T}
=\exp\biggl(\ep (T+V)+\,\ep^3e_{VTV} [V,[T,V]]+\cdots\biggr).
\label{th2b} 
\ee
For $v_1=0$, the first operator $\e^{t_1\ep T}$ classically
updates the position variable. Theorem 2b completely described the
structure of these {\it position}-type algorithms.
 
In both Theorem 2a and 2b,
one obtains fourth-order forward algorithms by simply moving the 
commutator $[V,[T,V]]$ term back to the left hand side and distribute it 
symmetrically among all the $V$ operators\cite{chin}. 

If some $t_i$ were allowed to be negative, then both $e_{TTV}$ 
and $e_{VTV}$ can be zero for $\delta g=0$. For both (\ref{vtvv}) 
and (\ref{ttvv}) we have 
\be
C_2=-\frac12
\ee
and  
\be
v_i=\frac12(t_i+t_{i+1}).
\ee
The latter is now true even for $i=1$ and $i=N$. 
This is not an coincident,
from (\ref{ttvf}) and (\ref{vtvf}), if we set $C_2=-1/2$, then
\be
e_{TTV}=2\, e_{VTV}=\frac1{12}-\frac{g}4=\frac1{12}\delta g.
\la{nege}
\ee
Since $C_2$ here is a true constant, $\{v_i\}$ is 
linearly related to $\{t_i\}$.
We can formulate this explicitly as a theorem for 
negative-coefficient factorization yielding truly linear algorithms:

\noindent
{\bf Theorem 3}: If
\be
v_i=\frac12 (t_i+t_{i+1})
\la{vif}
\ee
such that $t_1=0$, then
\be
\prod_{i=1}^N
\e^{t_i\ep T}\e^{v_i\ep V}
=\exp\biggl(\ep e_T(T+V)
+\,\frac{\ep^3}{24}\delta g\, \Bigl( 2 [T,[T,V]]+[V,[T,V]]\Bigr)+\cdots\biggr).
\label{prodlin} 
\ee
Both commutators now vanish simultaneously if $\delta g=0$.

An immediate corollary is that if $\delta g$ were to vanish, then there 
must be at least one $t_k<0$ such that $t_k^3+t_{k+1}^3<0$ {\it or }
$t_k^3+t_{k-1}^3<0$. Since 
$$(x^3+y^3)=(x+y)[\frac34 y^2+(x-\frac12 y)^2],$$
$x^3+y^3<0\Longrightarrow x+y<0$. We therefore must have $t_k+t_{k+1}<0$ or
$t_k+t_{k-1}<0$. From (\ref{vif}), this implies that $v_k$ or $v_{k-1}$ 
must be negative. Thus an algorithm of order greater than two
of the form (\ref{prodlin}) must contain at least one pair of negative
coefficients $t_i$ and $v_j$. In its general context, this is the 
Goldman-Kaper result\cite{goldman}. Our linear formulation here
is more precise: if $t_i$ is negative, then at least one of its 
adjacent $v_i$ must be negative. If only one $t_k$ is negative, 
then both of its adjacent $v_i$ must be negative.  

\section{The Structure of forward integrators}

Theorems 2a and 2b can be used to construct fourth-order forward 
algorithms with only positive factorization coefficients. These
forward integrators are the only fourth-order factorized symplectic
algorithms capable of integrating {\it time-irreversible} equation such as 
the Fokker-Planck\cite{fchinl,fchinm}
or the imaginary time Schr\"odinger equation\cite{auer,ochin,chinkro}.
Since it has been shown that\cite{nosix} currently there are no practical ways 
of constructing sixth-order forward integrators, these fourth-order 
algorithms enjoy a unique status.

For $N=3$, for a fourth-order algorithm, we must require
$t_2=t_3=1/2$. Theorem 2a then
implies that
\be
v_1=v_3=\frac16,\quad v_2=\frac23,\quad{\rm and}\quad e_{VTV}=-\frac1{72}.
\la{alg4a}
\ee
By moving the term $\ep^3 e_{VTV}[V,[T,V]]$ back to the LHS of (\ref{prod})
and combined it with the central $V$, one recovers
forward algorithm 4A\cite{suzfour,chin}.
For $N=4$ with $t_2=t_3=t_4=1/3$, we have
\be
v_1=v_4=\frac18,\quad v_2=v_3=\frac38,\quad{\rm and}\quad e_{VTV}=-\frac1{192}, 
\la{alg4d}
\ee
which corresponds to forward algorithm 4D\cite{chinchen01}. These are
special cases of the general {\it minimal} $|e_{VTV}|$, velocity-type 
algorithm given by
by  $t_1=0$, $t_i=1/(N-1)$, 
\be
v_1=v_N=\frac1{2N},\quad v_i=\frac{(N-1)}{N(N-2)},\quad{\rm with}\quad
e_{VTV}=-\frac1{24}\frac1{N(N-2)}.
\la{genvel}
\ee
This arbitrary $N$ algorithm can serve as a useful check for any
general fourth-order, velocity-type algorithm.

Alternatively, for $N=4$, we can allow $t_2$ to be a free parameter so that
\be
t_4=t_2,\quad t_3=1-2t_2.
\la{tbda}
\ee
Theorem 2a then fixes $C_2$ and $e_{VTV}$ with 
\be
\phi=6t_2(1-t_2)^2
\la{cbda}
\ee
and
\be
v_2=v_3=\frac1{12t_2(1-t_2)},\quad v_1=v_4=\frac12-v2
\la{vbda}
\ee
One recognizes that this is the one-parameter
algorithm 4BDA first found in Ref.\cite{chinchen02} using
symbolic algebra. For $t_2=1/2$, one
recovers integrator 4A; for $t_2=1/3$, one gets back 4D. The
advantage of using a variable $t_2$ is that one can use it to
minimize the resulting fourth-order error (oftentime to zero)
in any specific application. All the
seven-stage, forward integrators in the
velocity form described by Omelyan, Mryglod 
and Folk (OMF)\cite{ome03} correspond to different ways of
choosing $t_2$ and distributing the commutator term in 
4BDA. 

For $N=5$, again using $t_2$ as a parameter, 
we have $t_1=0$, $t_5=t_2$, $t_4=t_3=1/2-t_2$, (\ref{cande}) with
\be
\phi=\frac{15}{16}-3\left(t_2-\frac14\right)^2,
\ee
$v_5=v_1$, $v_4=v_2$, $v_3=1-2(v_1+v_2)$, and 
\be
v_1=\frac12+C_2(1-t_2),\quad v_2=-\frac12 C_2.
\la{veln5}
\ee 
This is a new one-parameter family of fourth order algorithms with 9 stages or
operators. 
 
To generate position-type algorithms, one can apply
Theorem 2b. 
For $N=3$, with
$v_1=0$, $v_1=v_2=1/2$, we have
\be
t_1=t_3=\frac12(1-\frac1{\sqrt{3}}),
\quad t_2=\frac1{\sqrt{3}},\quad{\rm and}\quad 
e_{VTV}=-\frac1{12}(1-\frac12\sqrt{3}).
\la{alg4b}
\ee
This produces forward algorithm 4B\cite{suzfour,chin} corresponding to 
$t_2=(1-1/\sqrt{3})/2$ in 4BDA. Again, this is a special case of the
general	fourth-order, minimal $|e_{VTV}|$ algorithm with
$v_1=0$, $v_i=1/(N-1)$,
\be
t_1=t_N=\frac12\left(1-\sqrt{\frac{N-2}N}\right),\quad 
t_i=\frac1{\sqrt{N(N-2)}},
\la{genpos}
\ee
and
\be
e_{VTV}=-\frac1{12} \left( 1-\frac{\sqrt{N(N-2)}}{(N-1)}\right).
\ee
For $N=4$, $v_1=0$ and
$v_2$ as the free parameter, invoking Theorem 2b gives
\be
v_4=v_2,\quad v_3=1-2v_2,
\ee
\be
t_2=t_3=\frac1{2\sqrt{6v_2}},\quad t_1=t_4=\frac12-t_2
\la{tacd}
\ee
and
\be
e_{VTV}=-\frac1{12}\left[1-(1-v_2)\sqrt{6v_2}\,\right]
\ee
For $v_2=1/6$ and $v_2=3/8$, this reproduces algorithm 4A and 4C\cite{chin}
respectively. One again recognizes that the above
is the one-parameter algorithm 4ACB first derived in Ref.\cite{chinchen02},
but now with a much simpler parametrization. Algorithm 4ACB covers all the
seven-stage, forward fourth-order position-type integrators
 described by OMF\cite{ome03}.  

For $N=5$, with $v_2$ as a free parameter, we have $v_1=0$, $v_5=v_2$,
$v_4=v_3=1/2-v_2$, and Theorem 2b produces another 9-stage fourth-order
algorithm with
\be
\phi^\prime=\sqrt{ {15}/{16}-3(v_2-1/4)^2 }.
\ee
$t_5=t_1$, $t_4=t_2$, $t_3=1-2(t_1+t_2)$, and 
\be
t_1=\frac12+C_2(1-v_2),\quad t_2=-\frac12 C_2.
\la{posn5}													
\ee 

For $N<5$, we have shown above that all fourth-order algorithms
are necessarily extended-linear. For $N\ge 5$, this is not necessary 
the case. Nevertheless we find that, remarkably, most known 
$N=5$ (9 stages) forward algorithms are very close to being
extended-linear. For velocity-type, 
$N=5$ extended-linear algorithms, 
$v_1$ and $v_2$ are functions of $t_2$ fixed by (\ref{veln5}).
In Fig.1, we compare this predicted relationship with the actual
values of $v_1$, $v_2$ and $t_2$ of five forward, velocity-type,
fourth-order algorithms found by OMF\cite{ome03}. 
These are their Eqs.(52)-(56),
with their $\theta$, $\vartheta$, $\lambda$ corresponds to $t_2$
$v_1$, and $v_2$ respectively. Four out their five algorithms,
with $v_1$ in particular, are well described by (\ref{veln5}).  

In Fig.2, we compare the coefficients of all three of OMF's
forward, position-type algorithms, Eq.(59)-(61),
with (\ref{posn5}), which fixes $t_1$, $t_2$ as a function of $v_2$.
Here, their parameters $\lambda$, $\rho$, $\theta$ correspond to
$v_2$, $t_1$, $t_2$ respectively. Again, $t_1$ is particularly well
predicted by (\ref{posn5}).

For 11-stage algorithms with $N=6$, we have two free parameters $t_2$, $t_3$
for velocity type algorithms with
\be
\phi=1-2t_2^3-2t_3^3-(1-2t_2-2t_3)^3
\ee
and two free parameters $v_2$, $v_3$ for position type algorithms
with
\be
\phi^\prime=\sqrt{1-2v_2^3-2v_3^3-(1-2v_2-2v_3)^3}.
\ee
Once $\phi$ and $\phi^\prime$ are known, we can determine $v_1$ and $v_2$
in the case of velocity-type algorithms and $t_1$ and $t_2$ in the case of
position-type algorithms. There is one 11-stage velocity algorithm with
positive coefficients found by OMF; their Eq.(68) with
$\rho(=t_2)=0.2029$, $\theta(=t_3)=0.1926$,
\be
\vartheta(=v_1)=0.0667,\quad{\rm and}\quad \lambda(=v_2)=0.2620.
\ee
The last two values are to be compare with the values
given by Theorem 2a below at the same values of $t_2$ and $t_3$,
\be
v_1=0.0848,\quad{\rm and}\quad v_2=0.2060.
\ee
For OMF's 11-stage, position-type algorithm Eq.(78), with
$\vartheta(=v_2)=0.1518$, $\lambda(=v_3)=0.2158$, 
\be
\rho(=t_1)=0.0642,\quad{\rm and}\quad \theta(=t_2)=0.1920,
\ee
For the same values of $v_2$ and $v_3$, Theorem 2b gives 
\be
t_1=0.0659,\quad{\rm and}\quad t_2=0.1881.
\ee
It is remarkable that these 11-stage, fourth-order algorithms
derived by complex symbolic algebra, remained very close to
the values predicted by our extended-linear algorithms.

\section{The Structure of non-forward integrators}

Theorem 3 can be used to derive two distinct families of non-forward,
fourth-order algorithms. Consider first the case of $N=4$. 
For $t_1=0$ with
symmetric coefficients $t_4=t_2$, the constriants
\be
2t_2+t_3=1
\ee
\be 2t_2^3+t_3^3=0
\ee
have unique solutions
\be
t_2={1\over{2-2^{1/3}}}\quad{\rm and}\quad t_3=-{{2^{1/3}}\over{2-2^{1/3}}}.
\ee
Eq.(\ref{vif}) then yields
\be
v_1=v_4={1\over 2}{{1}\over{2-2^{1/3}}},
\quad
v_2=v_3=-{1\over 2}{{(2^{1/3}-1)}\over{2-2^{1/3}}}.
\ee
One recognizes that we have just derived the well known fourth-order
Forest-Ruth integrator\cite{for90}. Note that there is complete symmetry
between $\{t_i\}$ and $\{v_i\}$. For position type algorithm, we simply
interchange the values of $t_i$ and $v_i$. 

There are no symmetric solutions for $N=5$, for the same reason that 
there are also no solutions for $N=3$. 
For $N=2k$, we have the general condition
\be
2\sum_{i=2}^kt_i+t_{k+1}=1,
\ee
\be
2\sum_{i=2}^kt_i^3+t_{k+1}^3=0,
\ee
which can be solved by introducing real parameters
$\alpha_i$ for $i=2$ to $k$ with $\alpha_2=1$,
\be
t_i=\alpha_i t_2,
\la{alphai}
\ee
so that
\be
t_{k+1}=-2^{1/3}\left( \sum_{i=2}^k\alpha_i^3 \right)^{1/3} t_2,
\ee
\be
t_2=\frac1{2\left(\sum_{i=2}^k\alpha_i\right)
-2^{1/3}\left( \sum_{i=2}^k\alpha_i^3 \right)^{1/3}}.
\ee
These solutions generalize the fourth-order Forest-Ruth integrator to
arbitrary $N$. 

For $N=2k+1$, $k>2$, again introducing (\ref{alphai})
for $i=2$ to $k$ with $\alpha_2=1$, we have
\be
t_{k+1}=-\left( \sum_{i=2}^k\alpha_i^3 \right)^{1/3} t_2,
\ee
\be
t_2=\frac1{2\left(\sum_{i=2}^k\alpha_i\right)
-2\left( \sum_{i=2}^k\alpha_i^3 \right)^{1/3}}.
\ee
This is a new class of fourth-order algorithm possible only
for $N$ odd and greater than five.

\section{Conclusions}

Most of the machinery for tracking coefficients were developed in
Ref.\cite{nosix} for the purpose of providing a constructive proof
of the Sheng-Suzuki theorem. The advantage of this constructive
approach is that we can obtain explicit lower bounds on the
the second-order error coefficients. Here, by imposing the 
extended-linear relationship between $\{t_i\}$ and $\{v_i\}$, 
these bounds become the actual error coefficients and provide
a complete characterization all fourth-order symplectic integrators 
for arbitrary number of operators.  
The most satisfying aspect of this work is that most fourth order 
integrators can now be derived analytically without recourse to 
symbolic algebra or numerical root-finding. We have also provided 
explicit construction of many new classes of fourth-order algorithms.

For $N=5,6$, corresponding to 9 and 11 operators, we have shown that 
many fourth-order algorithms found by Omelyan, Mryglod and Folk\cite{ome03} 
are surprisely close to the predicted coefficients of our theory, 
suggesting that the extended-linear relation between coefficients 
may be the dominate solution of the order-condition.  

The expansion (\ref{uexp}) may hold similar promise for characterizing
sixth order algorithms by introducing extended-quadratic or higher order 
relationships between the two sets of coefficients.

\begin{acknowledgments}

I thank Dr. Blanes and Casas for an interesting comment which led to this work. 
This work is supported, in part, by a National Science Foundation 
grant, No. DMS-0310580.

\end{acknowledgments}

\newpage
\centerline{REFERENCES}

\newpage
\begin{figure}
	\vspace{0.5truein}
	\centerline{\includegraphics[width=0.8\linewidth]{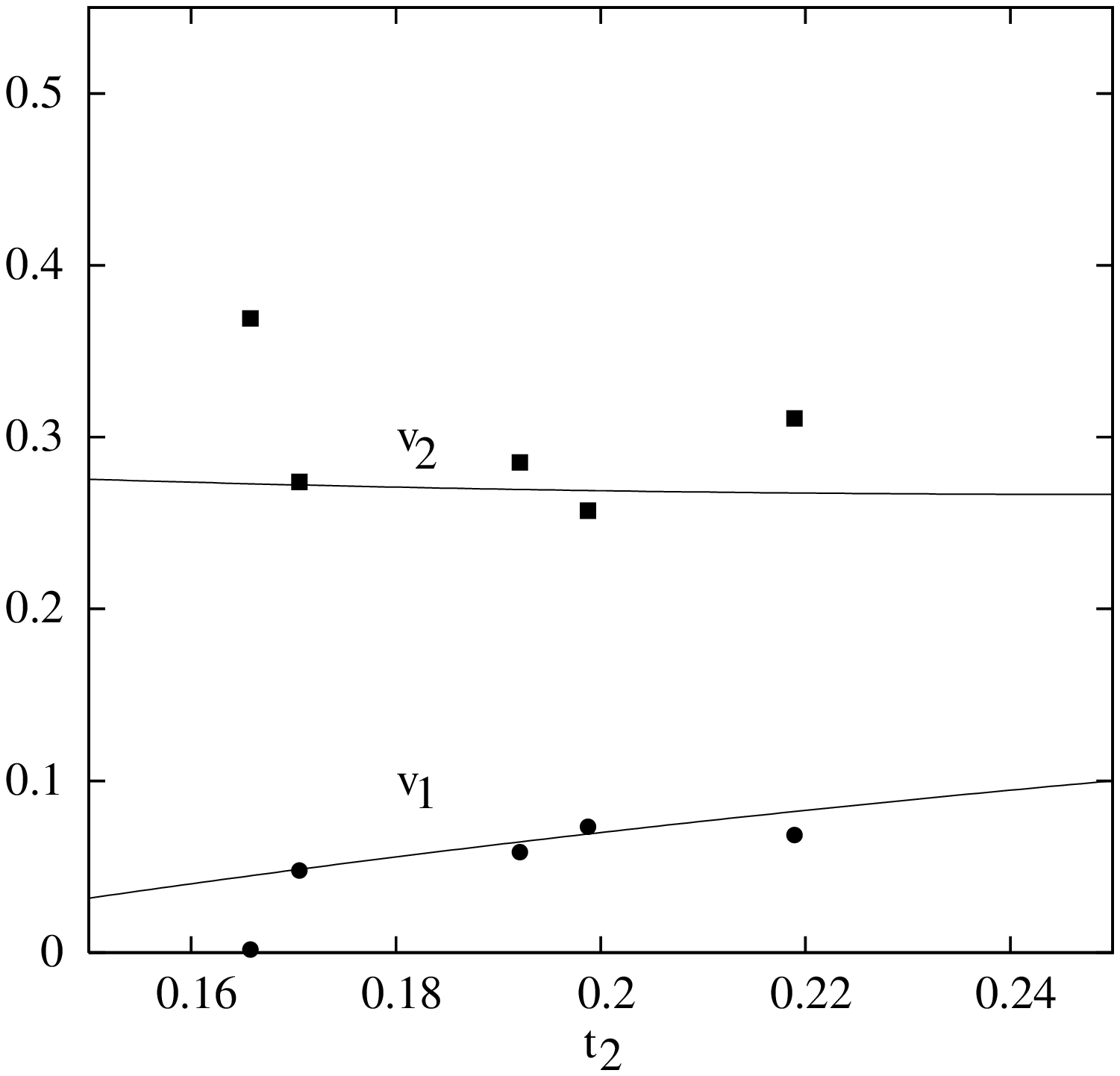}}
	\vspace{0.5truein}
\caption{Comparing the coefficients of five, 9-stage, {\it velocity}-type,
fourth-order forward integrators of Omelyan, Mryglod and Folk\cite{ome03} 
(filled circles and squares), with the analytical prediction of 
extended-linear symplectic integrators (solid lines). 
\label{fig1}}
\end{figure}
\newpage
\begin{figure}
	\vspace{0.5truein}
	\centerline{\includegraphics[width=0.8\linewidth]{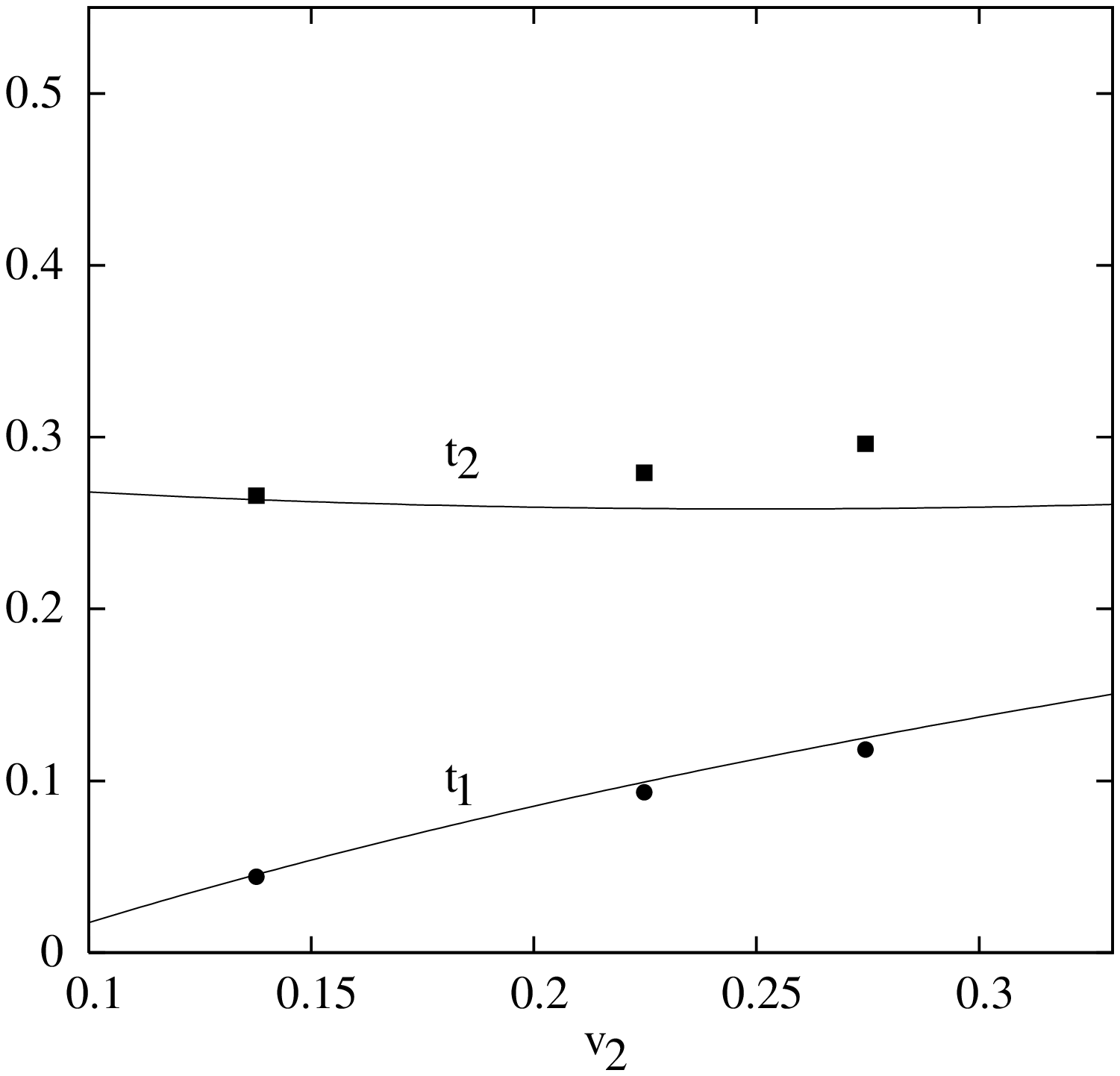}}
	\vspace{0.5truein}
\caption{
 Comparing the coefficients of three, 9-stage, {\it position}-type,
fourth-order forward integrators of Omelyan, Mryglod and Folk\cite{ome03} 
(filled circles and squares), with the analytical prediction of 
extended-linear symplectic integrators (solid lines).
\label{fig2}}
\end{figure}
\end{document}